\documentclass[a4paper,fleqn]{cas-dc}
\usepackage{natbib}
\usepackage{subfig}
\usepackage{graphicx}
\usepackage{balance}
\usepackage{tabularx}
\usepackage{caption}
\usepackage{hyperref}

\begin{document}
\shorttitle{Meteoric energy delivery on Titan}
\shortauthors{Flowers et al.}
\title[mode = title]{Energy Delivery via Meteors into Titan's Atmosphere}

\author[1]{Erin E. Flowers}[orcid=0000-0001-8045-1765]
\fnmark[1]
\cormark[1]
\address[1]{Department of Astrophysical Sciences, Princeton University, 4 Ivy Lane, Princeton, NJ 08544,USA}

\author[1, 2]{Christopher F. Chyba}
\address[2]{Princeton School of Public and International Affairs, 20 Prospect Ave, Princeton, NJ 08544, USA}

\author[3]{Paul J. Thomas}
\address[3]{Department of Physics and Astronomy, University of Wisconsin, 101 Roosevelt Avenue, Eau Claire, WI 54701, USA}

\cortext[1]{Corresponding author. \textit{E-mail address: \href{eflowers@princeton.edu}{eflowers@princeton.edu}}}
\fntext[1]{NSF GRFP Fellow}

\begin{abstract}
The Cassini-Huygens mission measured the chemical abundances of the major components of Titan's atmosphere, and analyses of the data revealed several as-yet unexplained anomalies in the methane and hydrogen profiles. We model the deceleration and ablation of meteors in Titan's atmosphere to examine whether meteor energy deposition could explain, in part, two of these anomalies. Our simulations vary meteor entry mass, trajectory angle, and velocity, and follow changes in all three as our meteors descend into a realistic Titan atmosphere. For the smallest particles, which deliver the most mass and therefore energy to Titan, we find that the altitudes where energy deposition peaks correspond to those of the observed chemical anomalies. In the region directly above the anomalies, energy deposition by meteors is greater than energy deposition from ultraviolet photons, which are typically responsible for methane dissociation. Finally, we calculate the total amount of energy available for chemical reactions in question. Total meteor energy deposited is swamped by daytime ultraviolet light, but of course is the dominant source of energy for atmospheric chemistry at the relevant altitudes during the night.

\end{abstract}


\maketitle

\section{Introduction}
What we know of Titan's atmosphere and environment comes largely from the Cassini-Huygens mission. During its two decades of operation, the Cassini spacecraft made over 100 close approaches to Titan, and the Huygens lander took over two hours of data during its descent through the moon's atmosphere. The atmospheric structure was determined with the Huygens Atmospheric Structure Instrument (HASI) \citep{Fulchignoni1997HASI}, while the composition was measured by the Gas Chromatograph Mass Spectrometer (GCMS) \citep{Niemann2002GCMS}. The dust environment of the Saturnian system was sampled by the Cosmic Dust Analyzer (CDA) \citep{Srama2004CDA} aboard the Cassini spacecraft, the first such instrument designed and deployed to specifically study the interplanetary and interstellar dust dynamics and composition in the vicinity of Saturn. Data on the meteoroid flux at Saturn has also been provided by the New Horizons and Pioneer 10 missions \citep{Stern2008NewHorizons, Horanyi2008SDC, Poppe2010SDCResults, Han2011EKB}.

These measurements confirmed earlier observations of the atmosphere's bulk properties, but they also revealed several anomalies in the atmospheric composition that remain unexplained \citep{Horst2017Review}. In this study, we explore the energy budget of Titan's atmosphere, specifically that portion provided by infalling meteors, to see if the energy provided by their deceleration and ablation in the atmosphere could be significant enough to explain the chemical anomalies observed by Cassini-Huygens, in particular, anomalies in the methane and hydrogen abundances (which will be described more in-depth in Section \ref{sec:results}).

Many studies that need to incorporate  the flux of interplanetary dust particles (IDPs) beyond the orbit of Earth employ the dust flux model of \citet{Grun1985Meteoric}. This model assumes that the isotropic dust flux of meteoroids measured at Earth's heliocentric distance may be used to extrapolate the dust flux throughout the solar system. Using this assumption, Grün {\em et al.} determine that the velocity of the interplanetary dust particles (IDPs) (which then is used to determine the flux) can be described by the relation:
\begin{equation}
    v_{\textnormal{x AU}} = \frac{v_{\textnormal{1 AU}}}{\sqrt{x}},
\end{equation}
as one would expect from simple scaling by Kepler's third law, where the velocity at a certain heliocentric distance $v_{\textnormal{x AU}}$ scales as the distance $x^{-1/2}$.
These assumptions overlook two potentially major effects when we consider meteoroids in the realm of the moons of the outer gas giants. The Grün {\em et al.} model is based on observations within 1 AU, which has a different radiation environment, as well as different gravitational effects due to the inner planets versus the outer. Moreover, the Grün {\em et al.} model's IDP flux is dominated by long-period comets, but in the outer solar system the particle population is dominated by the objects in the Edgeworth-Kuiper belt \citep{Stern1996KB, Yamamoto1998Dust}.

Nevertheless, the Pioneer spacecraft did tentatively detect a flux at Saturn that was a factor of $\sqrt{10}$ lower than the flux around 1 AU, and this has been cited as the reason for maintaining the Grün {\em et al.} model in subsequent studies at the Saturnian system \citep{Cuzzi1998MeteoroidEvolution}. But it has now been shown by \cite{Poppe2012EKB} that the Grün {\em et al.} model in fact both underestimates and overestimates the particle flux at Saturn by as much as two orders of magnitude depending on particle size. Therefore we choose to abandon the Grün {\em et al.} model. In Section \ref{sec:model}, we describe the particle-flux model we use instead.

In Section \ref{sec:model} we also present our models for meteor ablation and energy dissipation in Titan's atmosphere, including a discussion of the many ways our treatment differs from earlier work. The results of these models are reported in Section \ref{sec:results}, and then discussed in the context of their potential impact on Titan's atmospheric chemistry in Section \ref{sec:discuss}. Finally, we summarize our results and discussion points in Section \ref{sec:conclude}.

\section{Meteor Model}
\label{sec:model}
Previous modeling of meteoric entry into Titan's atmosphere has been based on the equations for meteor velocity, ablation, and thermal radiation given by \cite{Lebedinets1973Model} and then
propagated through the literature (e.g.  \cite{Pesnell2000Mars}, \cite{MolinaCuberos2001Ionization}, \cite{Pandya2014Sim}, \cite{Popova2019Modelling}).  However, these equations contain major errors that appear to have gone uncorrected, and that significantly affect the results of atmospheric entry simulations. We briefly summarize these errors and their effects here, before describing the model we employ.

The Lebedinets {\em et al.} ablation equation for a meteor of mass $m$ and density $\delta$ is:
\begin{equation}
    \frac{dm}{dt} = -\frac{4AC_1m^{2/3}}{\delta^{2/3}T^{1/2}}e^{-C_2/T} - \frac{\Lambda_S A\rho m^{2/3}v^3}{2Q\delta^{2/3}}
    \label{eq:mass}
\end{equation}
where $t$ is time, A is the shape factor, $T$ is the temperature, C$_1$ and C$_2$ are constants that describe the dependence of the evaporation rate on the temperature (given as $6.92\times10^{10}$ g cm$^{-2}$ and $5.78\times10^4$ K respectively), $\Lambda_S$ is the sputtering coefficient given by the equation $\Lambda_S(T) = 6\times10^{-6}\exp{(T_m/290)}$, $\rho$ is atmospheric density (a function of altitude, and therefore time, as the meteor descends), $v$ is the meteor's velocity, and $Q$ is the energy of evaporation of a stony meteor ($6\times10^{10}$ erg g$^{-1}$). The corresponding energy equation is (\cite{MolinaCuberos2001Ionization}):

\begin{equation}\label{eq:temp}
\begin{aligned}
\frac{dT}{dt} = \frac{4A\rho v^3}{8C\delta^{2/3}m^{1/3}}(\Lambda - \Lambda_S) - \frac{4A\sigma T^4}{C\delta^{2/3}m^{1/3}} \\ - \frac{4A C_1 Q}{C\delta^{2/3}T^{1/2}m^{1/3}}e^{-C_2/T}
\end{aligned}
\end{equation}
 where $\Lambda$ is the heat transfer coefficient, \textit{C} is the specific heat, and $\sigma$ the Stefan-Boltzmann constant. The coefficients $\Lambda$ and $\Lambda_S$ are unitless. 
 
 Equations \ref{eq:mass} and \ref{eq:temp} are dimensionally incorrect in both terms of Equation \ref{eq:mass} and in the second and third terms of Equation \ref{eq:temp}. Instead of producing units of mass per second as required, the first term of Equation \ref{eq:mass} has units \textit{g s$^{-1}$ K$^{-1/2}$} and its second term has units of $s^{-1}$. The second and third terms in Equation \ref{eq:temp}, which should have units of temperature per second, are instead \textit{K s$^{-1}$ g$^{-1}$} and \textit{K$^{1/2}$ s$^{-1}$} respectively. The terms containing $T^{-1/2}e^{-C_2/T}$, which describe how ablation is affecting the mass and temperature, are particularly problematic. Evaluating these terms individually results in a value that is essentially zero, negating the effects of evaporation in the mass loss equation and loss of heat through ablation in the change in temperature equation. In our model, which uses corrected forms of these equations, we find the corresponding terms (in Equations \ref{eq:true_temp} and \ref{eq:true_mass} below) to be non-negligible. When comparing the terms in the temperature equations that pertain to ablation, we find that the ablation term in the Lebedinets formulation is 10$^{67}$ times smaller than the corresponding term in our equation \ref{eq:true_temp}. As such, we caution against the continued use of the meteor entry equations from \cite{Lebedinets1973Model}, or results from papers that have directly adopted these equations.

To avoid these problems, we instead use the corresponding, but dimensionally correct and physically justified, equations for energy, velocity, and ablation described in \cite{CampbellBrown2004Model} and \cite{Bronshten1983Physics} for meteoroid entry in Earth's atmosphere, with appropriate adaptation to Titan, and converting from a time step to an altitude step. The physical parameters adopted by \cite{CampbellBrown2004Model} are for Earth-atmosphere meteors of cometary origin, but even comet-originating dust at 1 AU will have been substantially devolatilized. We therefore adopt physical parameters from \cite{Ip1990Model} for icy meteors in the Saturnian system.

The energy equation is
 \citep{CampbellBrown2004Model}:
\begin{equation}\label{eq:true_temp}
    \begin{aligned}
        \frac{dT}{dz} = \frac{1}{Cmv\cos{\theta}}\Big[\frac{\Lambda\rho v^3}{2}A\Big(\frac{m}{\delta}\Big)^{2/3} \\
        - 4\sigma\epsilon\Big(T^4-T_a^4\Big)A\Big(\frac{m}{\delta}\Big)^{2/3} - L\frac{dm}{dz}\Big]
    \end{aligned}
\end{equation} 
where $\theta$ is the angle of the meteor's trajectory with respect to the vertical, $\epsilon$ is the emissivity of the meteor, $T_a$ is the temperature of the atmosphere at each altitude step, $L$ is the heat of ablation of an icy meteor, and the remaining variables are the same as in Equations \ref{eq:mass} and \ref{eq:temp}. The velocity equation is:
\begin{equation}\label{true_vel}
    \frac{dv}{dz} = \frac{1}{cos\theta}\frac{\Gamma\rho v}{m}A\Big(\frac{m}{\delta}\Big)^{2/3} - \frac{g}{v}
\end{equation}
where $g = g_T \big(\frac{R_T}{R_T + z}\big)^2$ ($g_T$ is the gravitational acceleration of Titan at surface level and $R_T$ is the radius of Titan) and $\Gamma$ is the drag coefficient, of order unity. Finally, the ablation equation is (\cite{Bronshten1983Physics}):
\begin{equation}\label{eq:true_mass}
    \frac{dm}{dz} = -\frac{1}{ cos\theta}\frac{A\Lambda}{2L}\Big(\frac{m}{\delta}\Big)^{2/3}\rho v^2
\end{equation}
Values of the constants appropriate for icy meteors are reported in Table \ref{tab:constants}.

\begin{table}[]
    \centering
    \begin{tabular}{m{0.15\linewidth}|m{0.35\linewidth}|m{0.35\linewidth}}
         \textbf{Variable} & \textbf{Value} & \textbf{Reference} \\
         \textit{C} & 4.18 $\times 10^7$ erg/g/K & \cite{Ip1990Model} \\
         $\Lambda$ & 1 & \cite{Briani2013Model} \\
         $\rho$ & 1 g/cm$^3$ & - \\
         $\epsilon$ & 0.9 & \cite{CampbellBrown2004Model} \\
         \textit{A} & 1.2 & \cite{Bronshten1983Physics} \\
         \textit{L} & 2.8 $\times 10^{10}$ erg/g & \cite{Ip1990Model} \\
         $\Gamma$ & 1 & \cite{CampbellBrown2004Model} \\
        
         $C_L$ & 0 & \cite{Love1994CL}
    \end{tabular}
    \caption{Values for constants, with references.}
    \label{tab:constants}
\end{table}

\begin{figure}
     \includegraphics[height=2in]{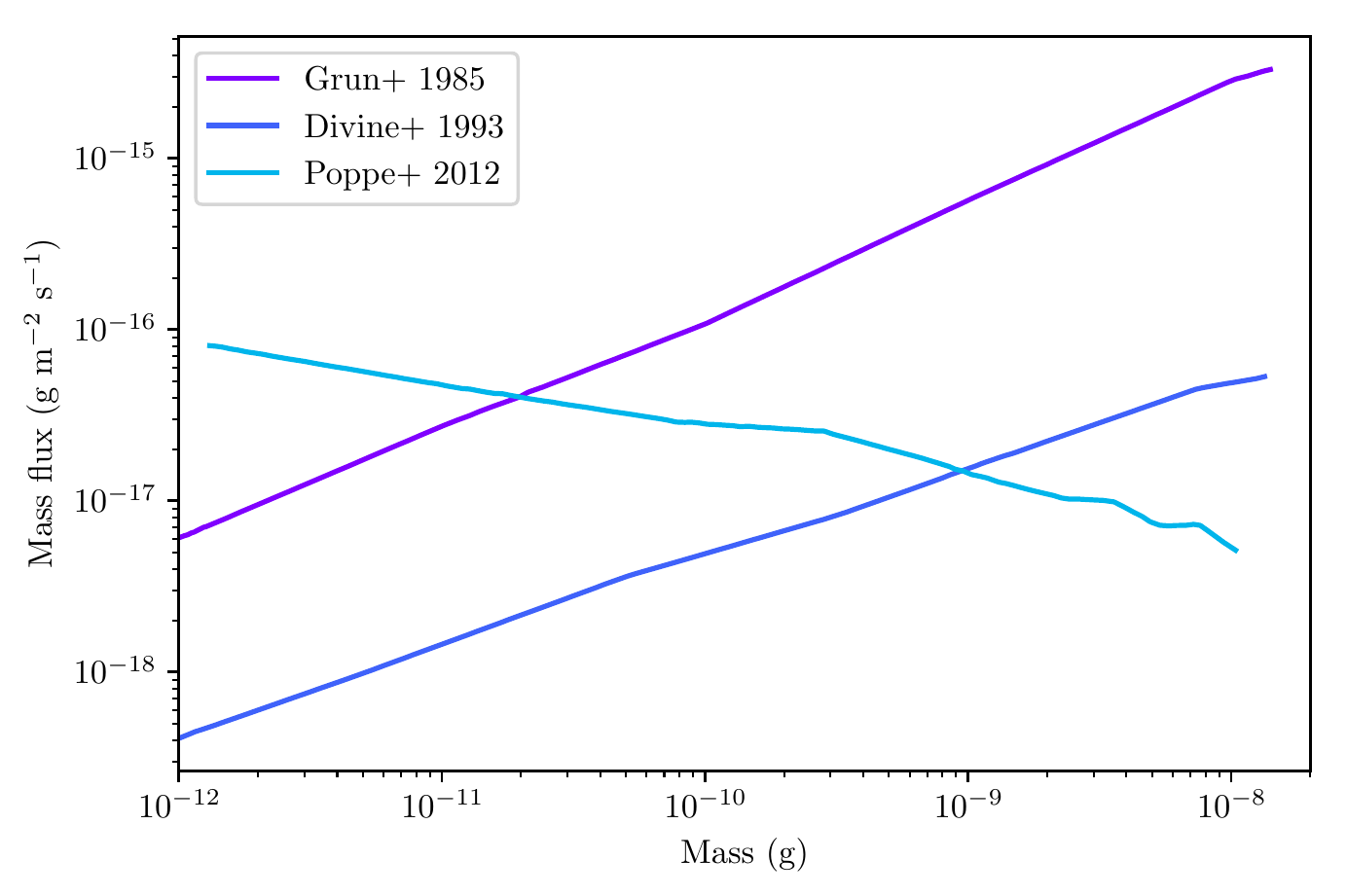}
     \caption{Mass flux at Saturn as a function of meteoroid mass, showing the flux models of \cite{Poppe2012EKB}, \cite{Grun1985Meteoric}, and \cite{Divine1993Model} for comparison. In the Grün and Divine models, which extrapolate from fluxes observed at 1 AU, the peak mass flux is for the largest (10$^{-8}$ g) particles, whereas in the \cite{Poppe2012EKB} model, based on spacecraft data in the outer Solar System, the peak mass flux is for the smallest (10$^{-12}$ g)
     particles, resulting in a very different energy availability within the atmosphere. We use the mass flux values from \cite{Poppe2012EKB} in our analysis.}
     \label{fig:mass_flux}
 \end{figure}

We solve these equations for a meteor moving from the top of the atmosphere to the surface using the RK4 method. We examine four cases that allow us to thoroughly explore the relevant parameter space: (1) using a constant trajectory angle (45$^{\circ}$ entry, which is the most probable entry angle for meteorites \citep{Hughes1993Angles}), we vary the initial mass of the meteor over the range 10$^{-12}$ $g$ -- $10^3 g$; (2) using a constant mass (10$^{-12}$ g), we vary the entry angle from 0 to 90$^{\circ}$; (3) holding the entry angle constant at 45$^{\circ}$, we vary the initial mass from 10$^{-12}$ $g$ -- 10$^3 g$ while solving for the changing trajectory angle at each altitude step using Equation \ref{eq:angle} \citep{Chyba1993Tunguska}:
\begin{equation}
\label{eq:angle}
\frac{d\theta}{dz} = \Big(\frac{1}{v\cos{\theta}}\Big) \Big(-\frac{g\sin{\theta}}{v} + \frac{C_L \rho S_A v}{2m} + \frac{v\sin{\theta}}{R_T + z}\Big)
\end{equation}
 where $C_L$ is the lift coefficient and $S_A$ is the cross-sectional area of the particle; and (4) once again using a constant mass (10$^{-12}$ g) and changing the initial entry angle from 0 to 90$^{\circ}$ while varying its trajectory angle at each altitude step. Regarding the lift coefficient, even for macroscopic meteors, $C_L < 10^{-3}$ \citep{Passey1980Meteors}, but it should be entirely negligible for micrometeors \citep{Love1994CL}. It is easy to show in Eq. \ref{eq:angle} that for $C_L < 10^{-3}$, the lift term is negligible for micrometeors in Titan’s atmosphere compared with the curvature term, and as such, we set $C_L = 0$. The constant mass is chosen to be 10$^{-12}$ g because this is the particle mass at which the mass flux of the observed meteoroid population at Saturn peaks (see Figure \ref{fig:mass_flux}, which we derived from data in Figure 2 of \cite{Poppe2012EKB}). Particles of this mass therefore should deliver the most energy into Titan's atmosphere. With respect to meteor  trajectory, previous Titan studies have typically assumed that the meteor maintains a straight-line path through the atmosphere; this is another way in which we deviate from previous work. For the atmospheric density as a function of altitude, we interpolate using data from HASI \citep{Fulchignoni2005HASIData}, which covers altitudes from 1500 km to the surface.
 
 In order to determine the amount of energy deposited by these micrometeors, we use the dust flux model of \citet{Poppe2012EKB}, given current lack of published flux values from the CDA. \citeauthor{Poppe2012EKB} use IDP observations from the New Horizons Student Dust Counter (SDC) and Pioneer 10 \citep{Stern2008NewHorizons, Horanyi2008SDC, Poppe2010SDCResults, Han2011EKB} to constrain a new dust flux model that takes into account the dynamics of the outer solar system's main source of dust - the Edgeworth-Kuiper Belt (EKB). Using the dynamical dust grain tracing model described in \cite{Han2011EKB}, \citeauthor{Poppe2012EKB} determine a flux at Saturn that departs by orders of magnitude from the flux determined by either the Grün {\em et al.} model or the \cite{Divine1993Model} model. We convert these models from number flux to mass flux and display the results in Figure \ref{fig:mass_flux}. 
 
 To include  gravitational focusing effects on Titan due to Saturn in their calculation of fluxes within the Saturnian system, \citet{Poppe2012EKB} use the equation:
 \begin{equation}
     G^a(r) = 1 + \frac{1}{2}\Big(\frac{v_{esc}(r)}{v^a_\infty}\Big)^2
 \end{equation}
 where $G^a(r)$ is the gravitational focusing factor as a function of distance from Saturn, $v_{esc}(r)$ is the Saturnian escape velocity (also a function of distance from Saturn), and $v^a_\infty$ is the grain impact velocity at infinity \citep{Colwell1994Disruption}. There is an additional focusing effect due to Titan's own gravity, but it is easy to show that this enhances the flux incident on Titan by $<10\%$, an amount lost in the other uncertainties in the problem. 
 
 \citeauthor{Poppe2012EKB} determine impact velocities within the Hill sphere of Saturn and, as a function of particle size, the impact flux. In order to then determine meteor impact velocities at Titan, one must also account for Titan's motion relative to Saturn. When we account for Titan's escape velocity, Saturn's escape velocity, and the impact speed at the Hill sphere radius, we calculate an average initial impact velocity to be $\sim$ 8 km/s (which is in agreement with \cite{Poppe2012EKB} impact velocities), though we do test a range of velocities between 2 km/s and 18 km/s for the four smallest particle radii to determine if this has a significant effect on their lifetime and energy deposition in the atmosphere, the results of which are reported in Section \ref{sec:vel}. At the orbit of Titan, the impact flux is order 10$^{-4}$ m$^{-2}$ s$^{-1}$ for the smallest particles and  10$^{-8}$ m$^{-2}$ s$^{-1}$ for the largest. While \citeauthor{Poppe2012EKB} confirm that these fluxes are constrained by SDC and Pioneer observations, they acknowledged at the end of their analysis that data from the Cosmic Dust Analyzer would be the best constraint. We use these values in our energy deposition determinations.

\section{Results}
\label{sec:results}
We are interested in understanding the energy budgets of regions of Titan's atmosphere where there have been observed chemical anomalies. \citet{Strobel2010Hydrogen} determines that, given the mixing ratios of H$_2$ in the lower thermosphere and the tropopause, there is a downward flux of H$_2$. Measurements of the H$_2$ mole fraction from the Cassini Ion Neutral Mass Spectrometer (INMS) (which probes the thermosphere) and the Cassini Composite Infrared Spectrometer (CIRS) (which probes the troposphere) differ by a factor $\gtrsim$ 2. In order to explain this observed profile, there must be a downward flux of H$_2$, and/or additional methane production. There are also anomalies in the methane abundance between 950 to 1500 km; although the mixing ratio is largely constant, around this altitude range it begins to vary \citep{MullerWodard2008Structure, Cui2009INMS, Magee2009INMS}. INMS measurements reveal a variation of the CH$_4$ mixing ration from 1.31\% at 81 km to 3\% at 1150 km that is not completely explained, in addition to latitudinal variations. The meteorite flux is relevant in understanding these anomalies because the infalling material is composed predominantly of volatile ices. These ices can provide material that increase the overall abundance of hydrogen-species in the atmosphere and provide energy that drives forward chemical reactions. 

\begin{figure*}
    \centering
    \subfloat[]{\includegraphics[height=2in, keepaspectratio]{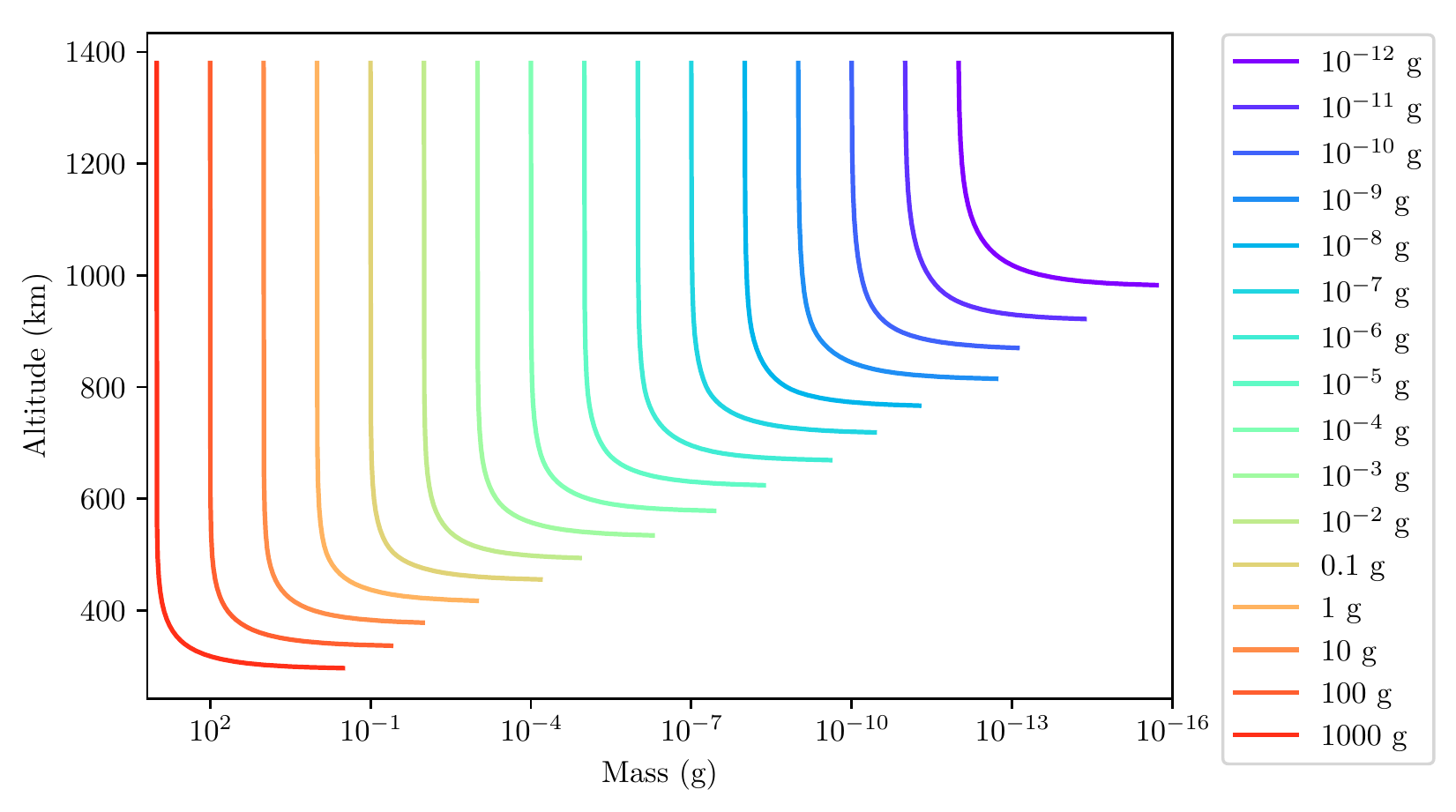}} \subfloat[]{\includegraphics[height=2in, keepaspectratio]{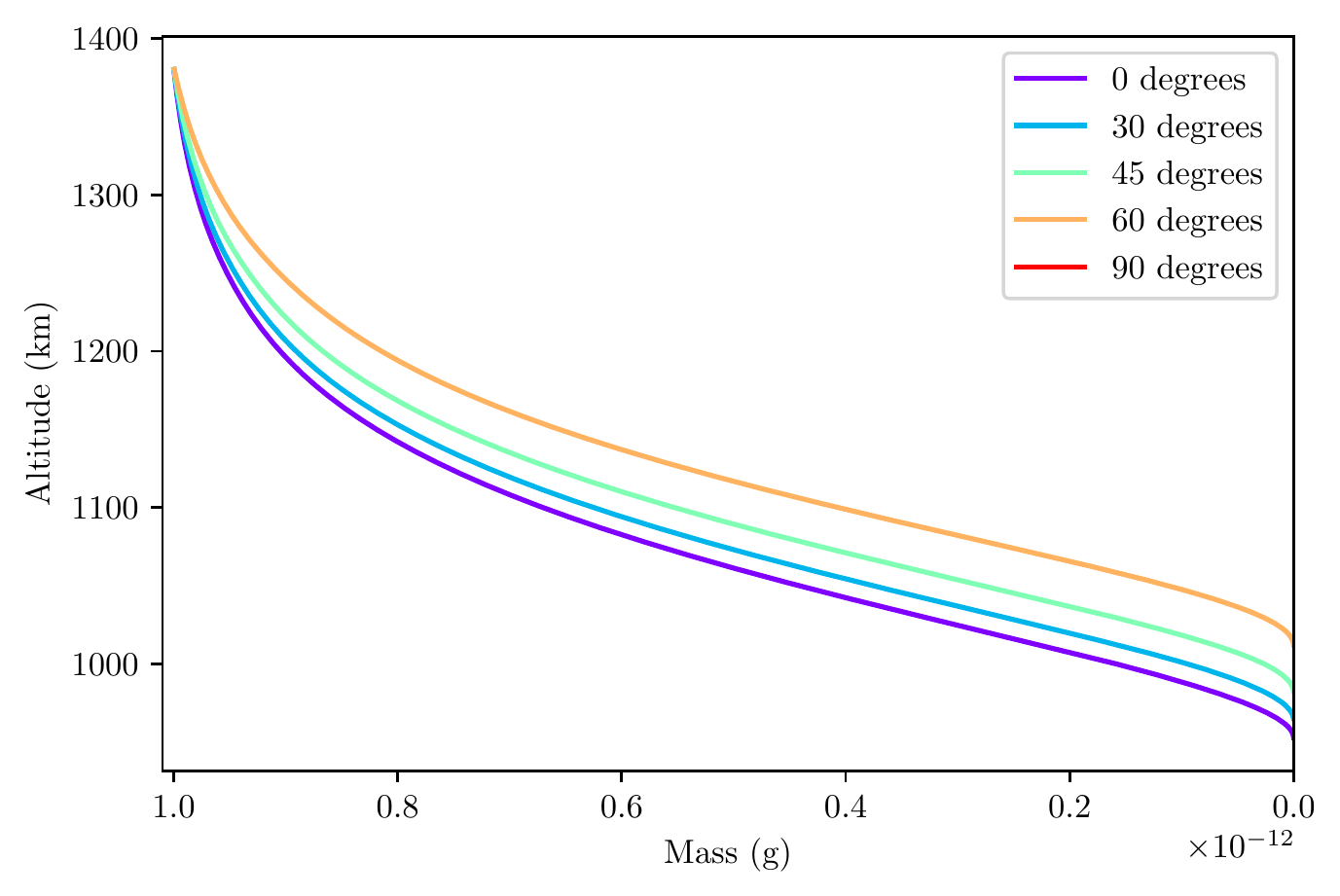}}
    \hspace{0mm}
    \subfloat[]{\includegraphics[height=2in, keepaspectratio]{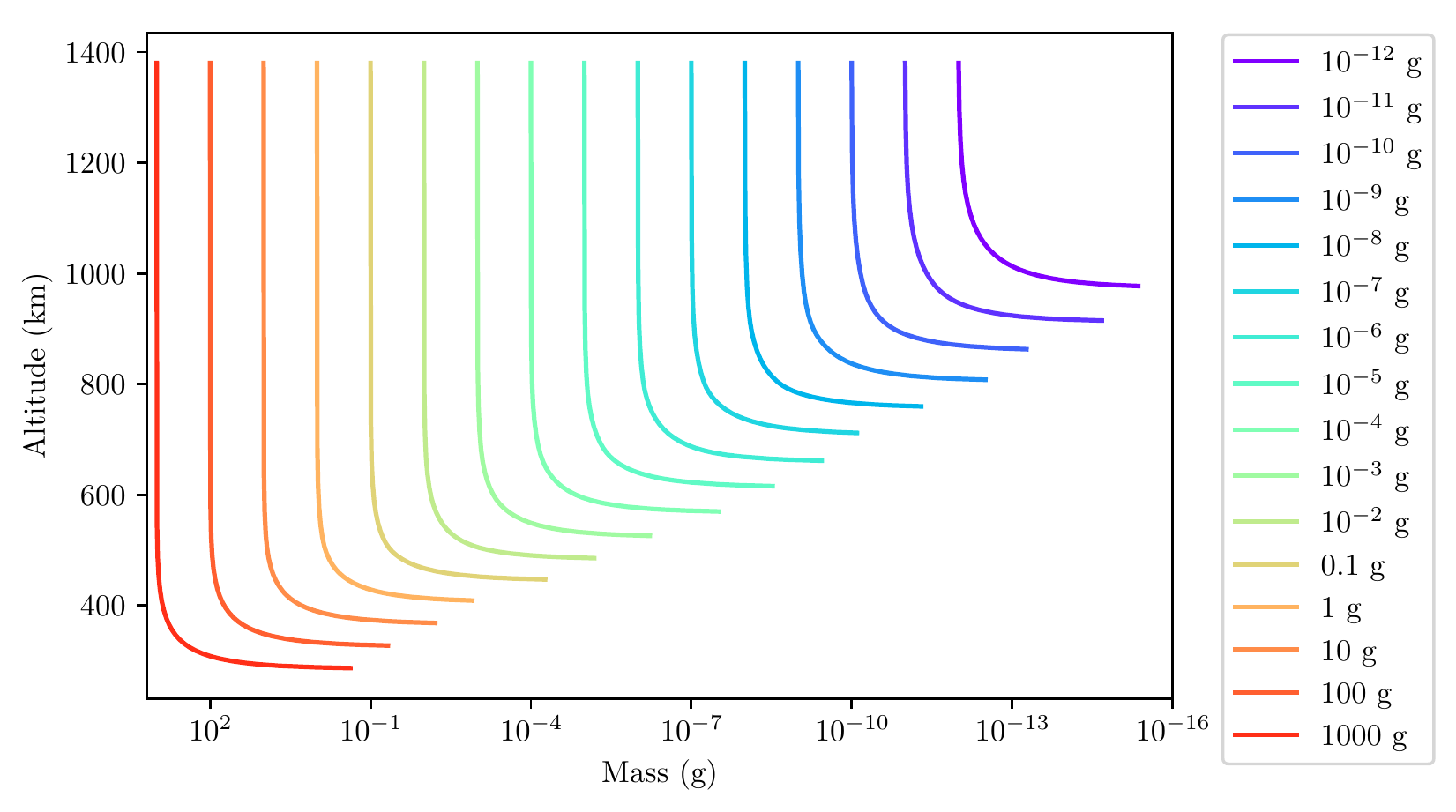}} \subfloat[]{\includegraphics[height=2in, keepaspectratio]{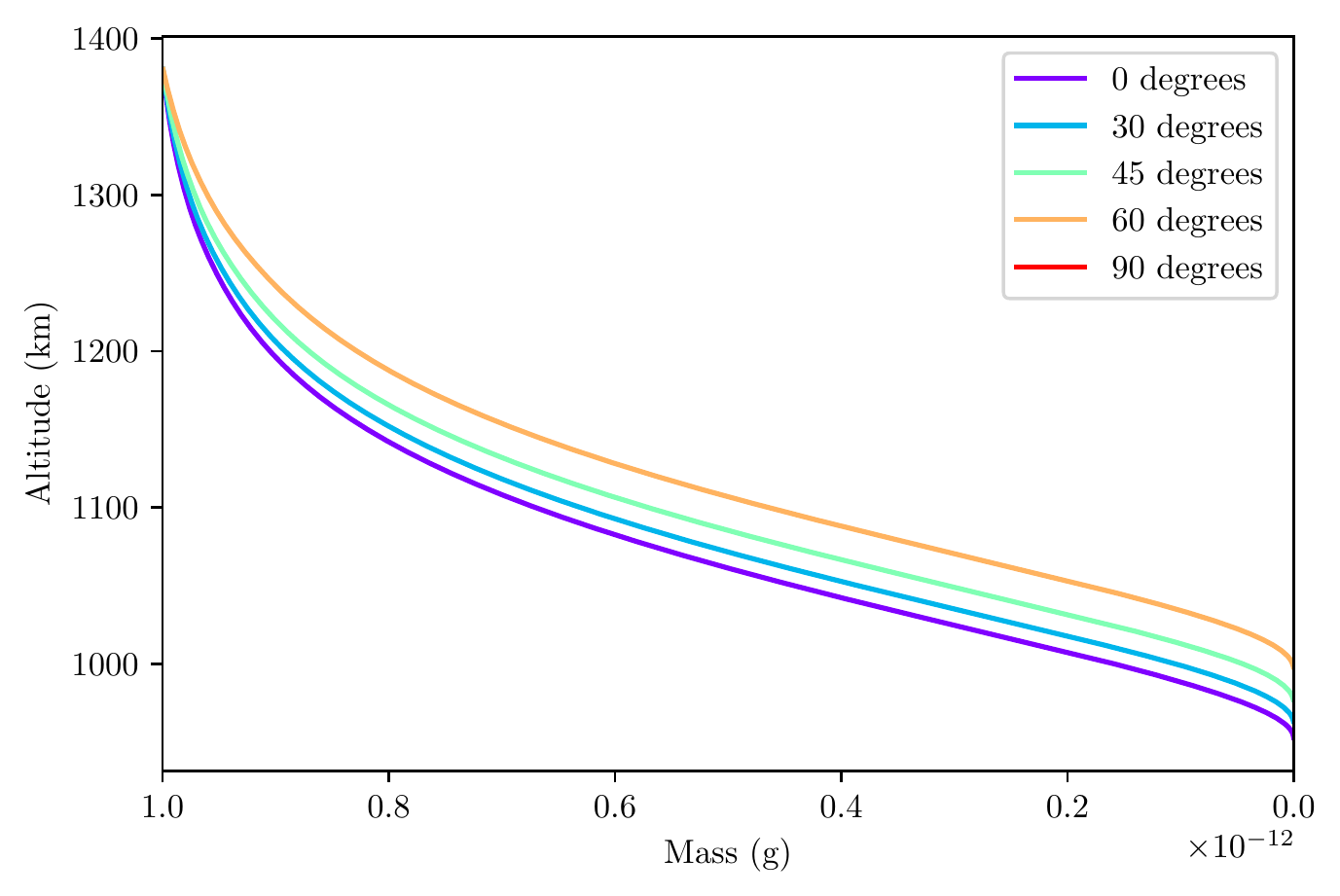}}
    \caption{(a) Particle mass as a function of altitude for a range of initial masses, for the case where the entry angle is 45$^\circ$ and the trajectory angle $\theta$ is held constant; (b) mass as a function of altitude for a range of initial entry angles, for the case where the initial mass is $10^{-12}$ g, and $\theta$ is held constant; (c) mass as a function of altitude for a range of  initial masses, the entry angle is 45$^\circ$, and $\theta$ varies over the course of the particle's descent according to 
    Equation \ref{eq:angle}; (d) mass as a function of altitude for a range of entry angles, where the initial mass is $10^{-12}$ g, and $\theta$ varies over the course of the particle's descent.}
    \label{fig:mass_alt}
\end{figure*}

\begin{figure*}
    \centering
    \subfloat[]{\includegraphics[height=2in, keepaspectratio]{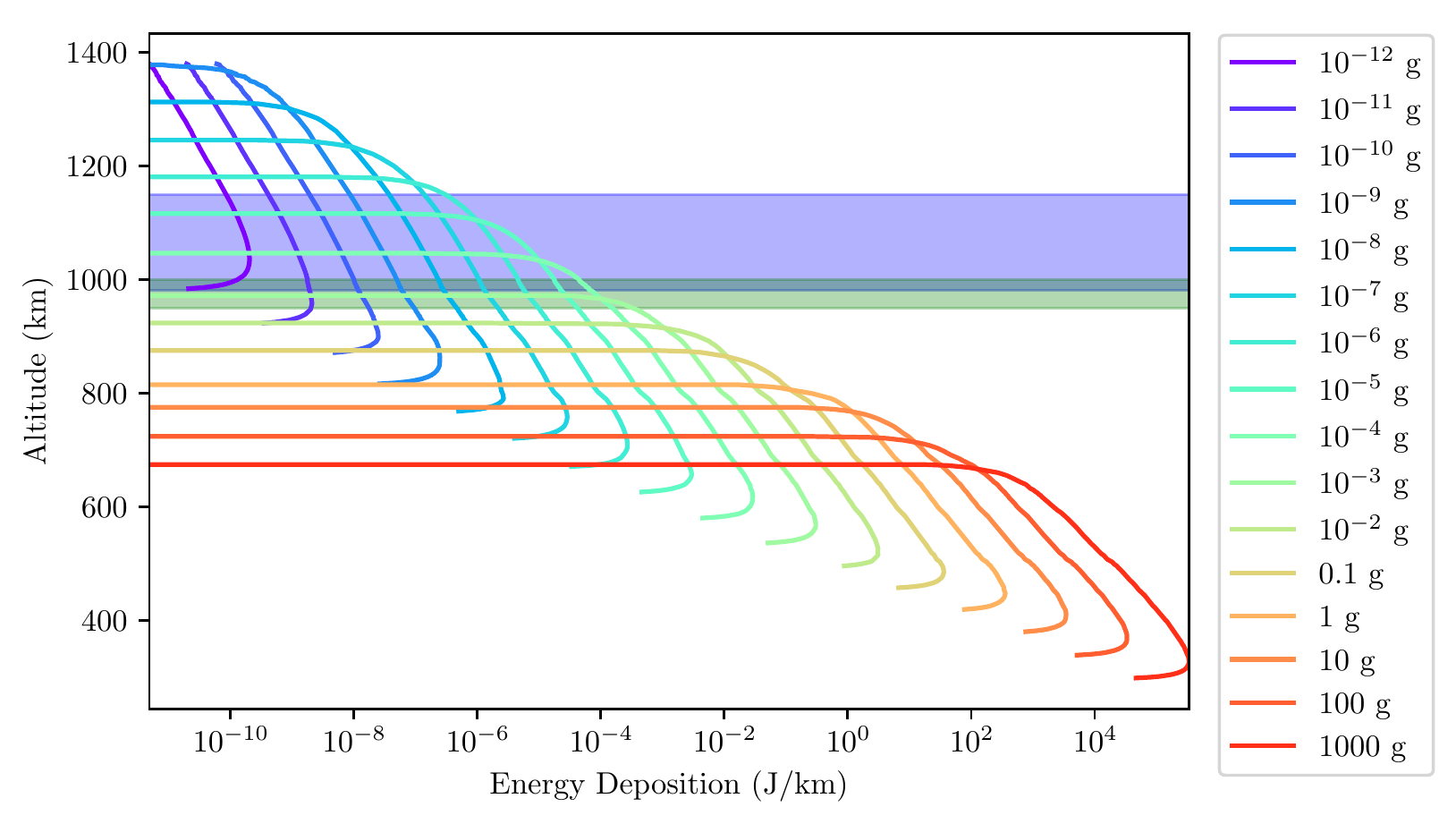}}
    \subfloat[]{\includegraphics[height=2in, keepaspectratio]{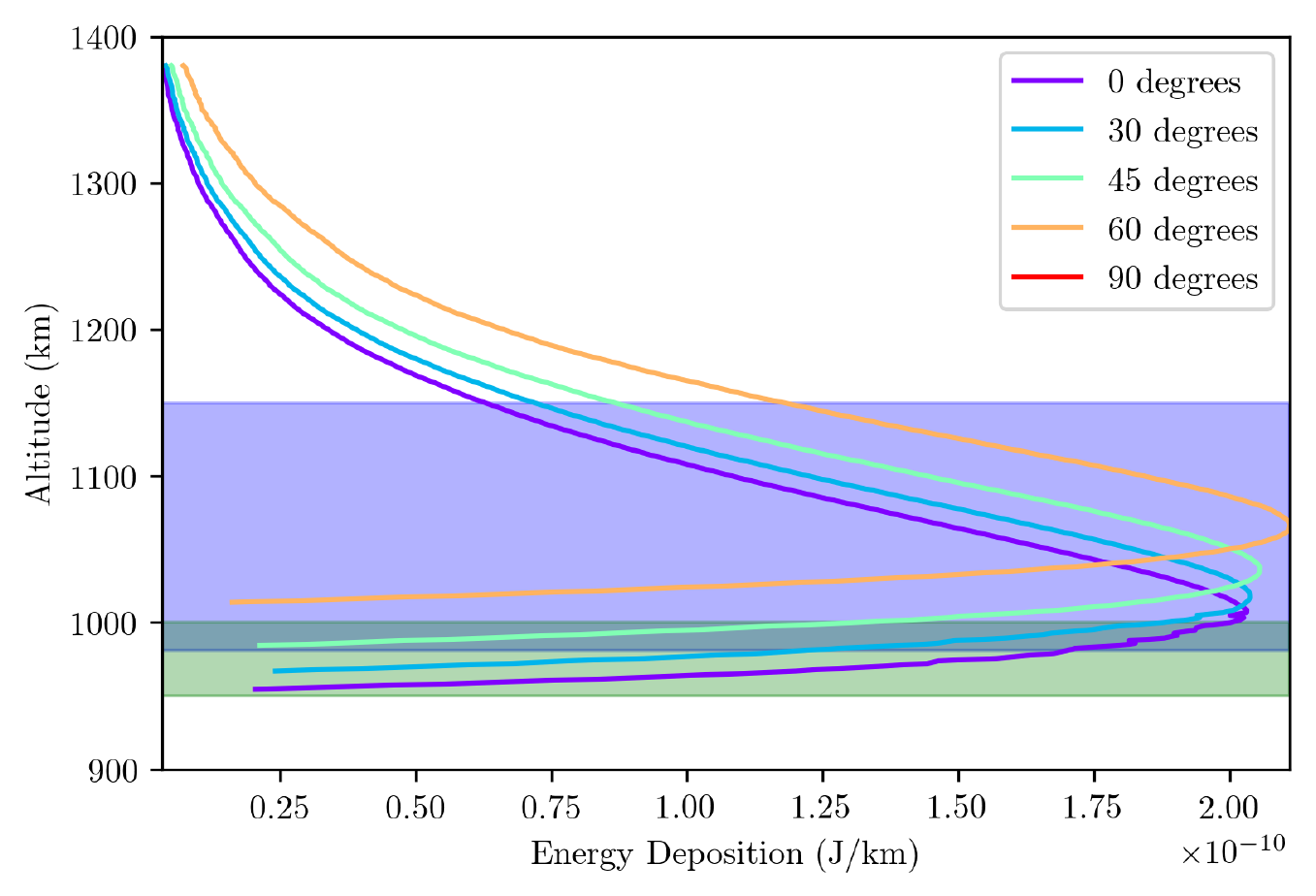}}
    \hspace{0mm}
    \subfloat[]{\includegraphics[height=2in, keepaspectratio]{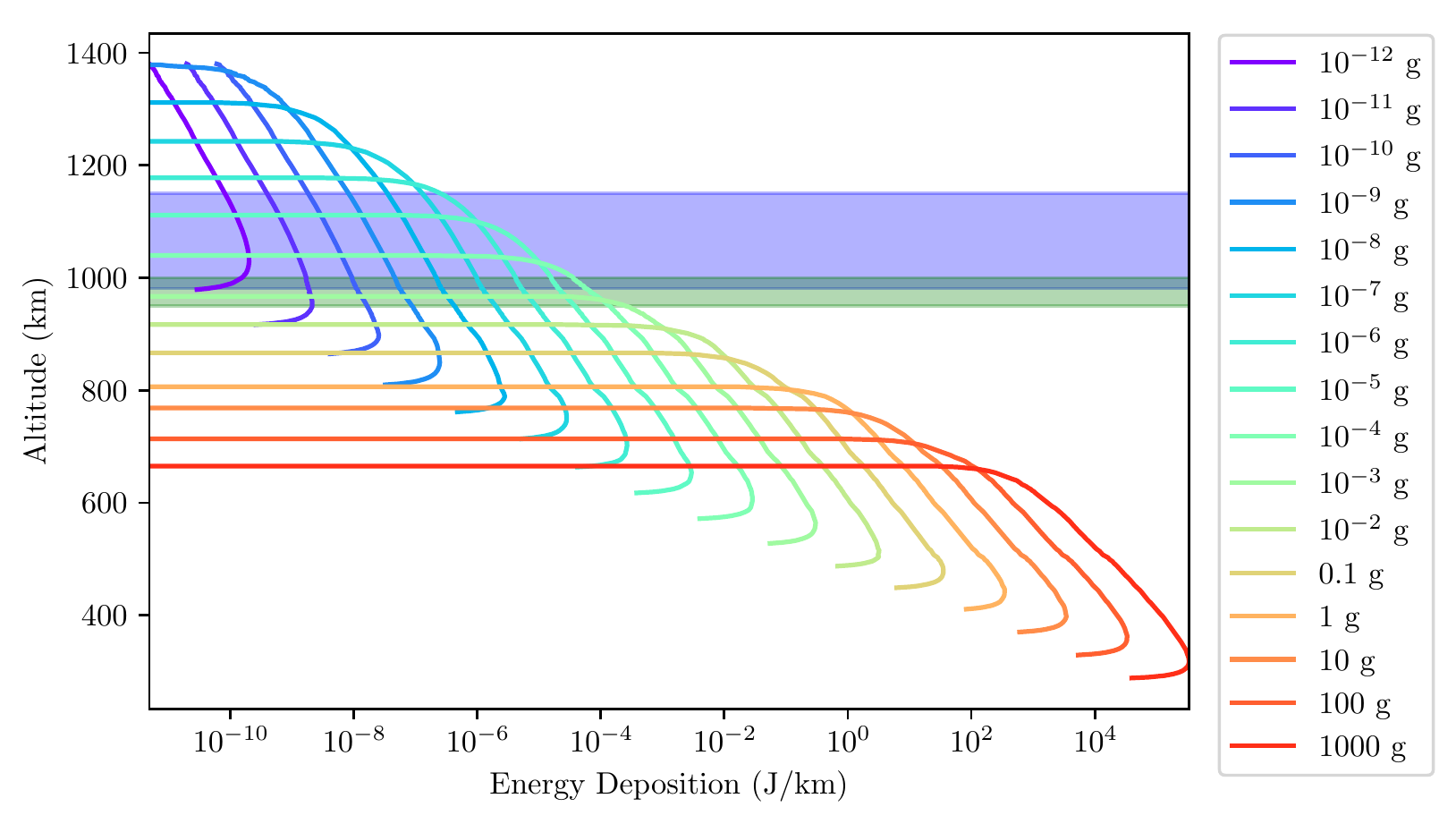}}
    \subfloat[]{\includegraphics[height=2in, keepaspectratio]{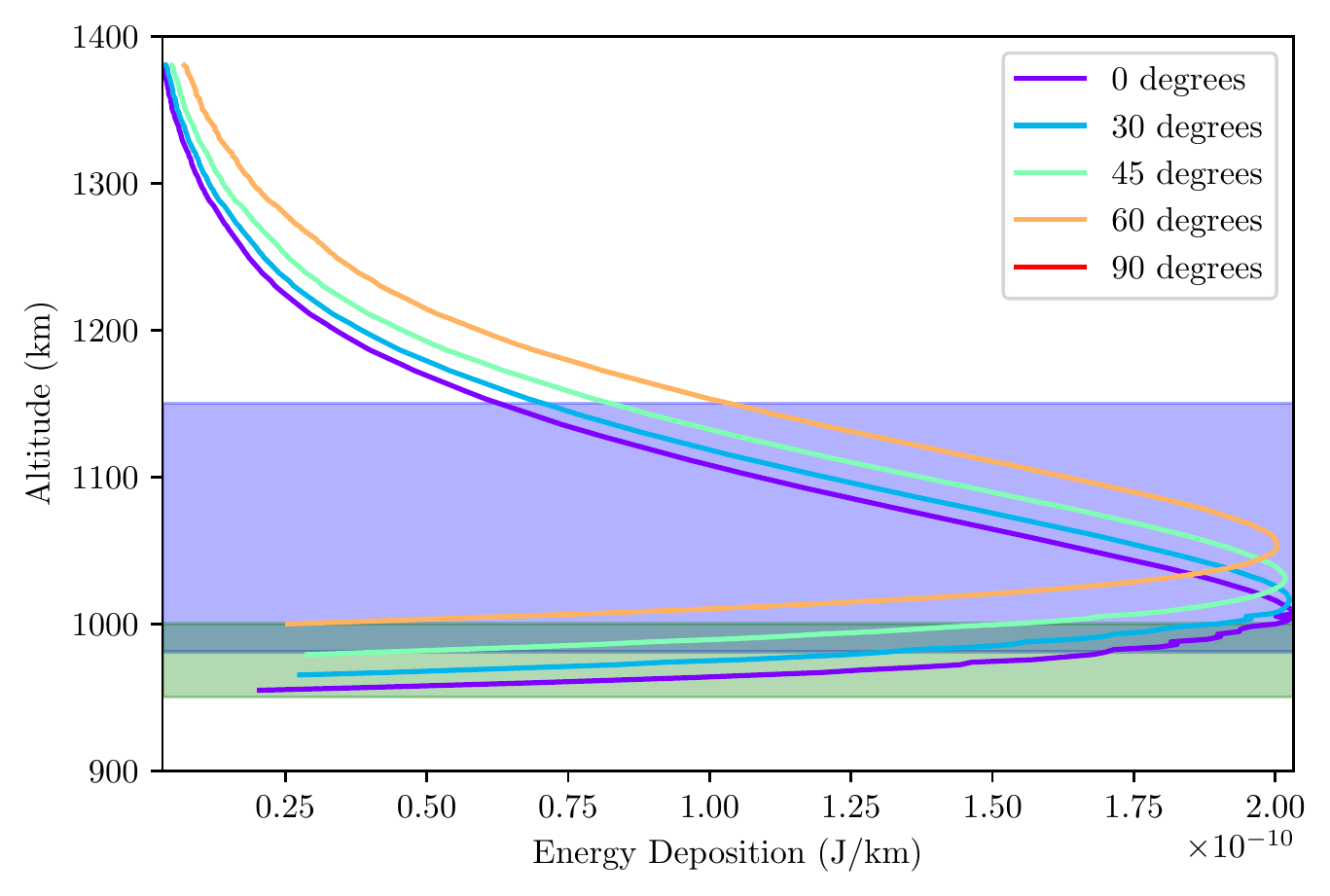}}
    \caption{In all four plots, the blue shaded region corresponds to the methane anomaly and the green shaded regions correspond to the molecular hydrogen anomaly as described in Section \ref{sec:results}. (a) Kinetic energy deposited as a function of altitude for the case where the initial mass is varied, the entry angle is 45$^\circ$ while the angle is held constant; (b) Kinetic energy deposited as a function of altitude for the case where the initial mass is $10^{-12}$ g, the entry angle is varied, and the angle is held constant; (c) Kinetic energy deposited as a function of altitude where the initial mass is varied, the entry angle is 45$^\circ$, and the angle varies over the course of the particle's descent; (d) Kinetic energy deposited as a function of altitude where the initial mass is $10^{-12}$ g, the entry angle is varied, and the angle varies over the course of the particle's descent.}
    \label{fig:kin_eng_dep}
\end{figure*}

\begin{figure*}
    \centering
    \subfloat[]{\includegraphics[height=2in, keepaspectratio]{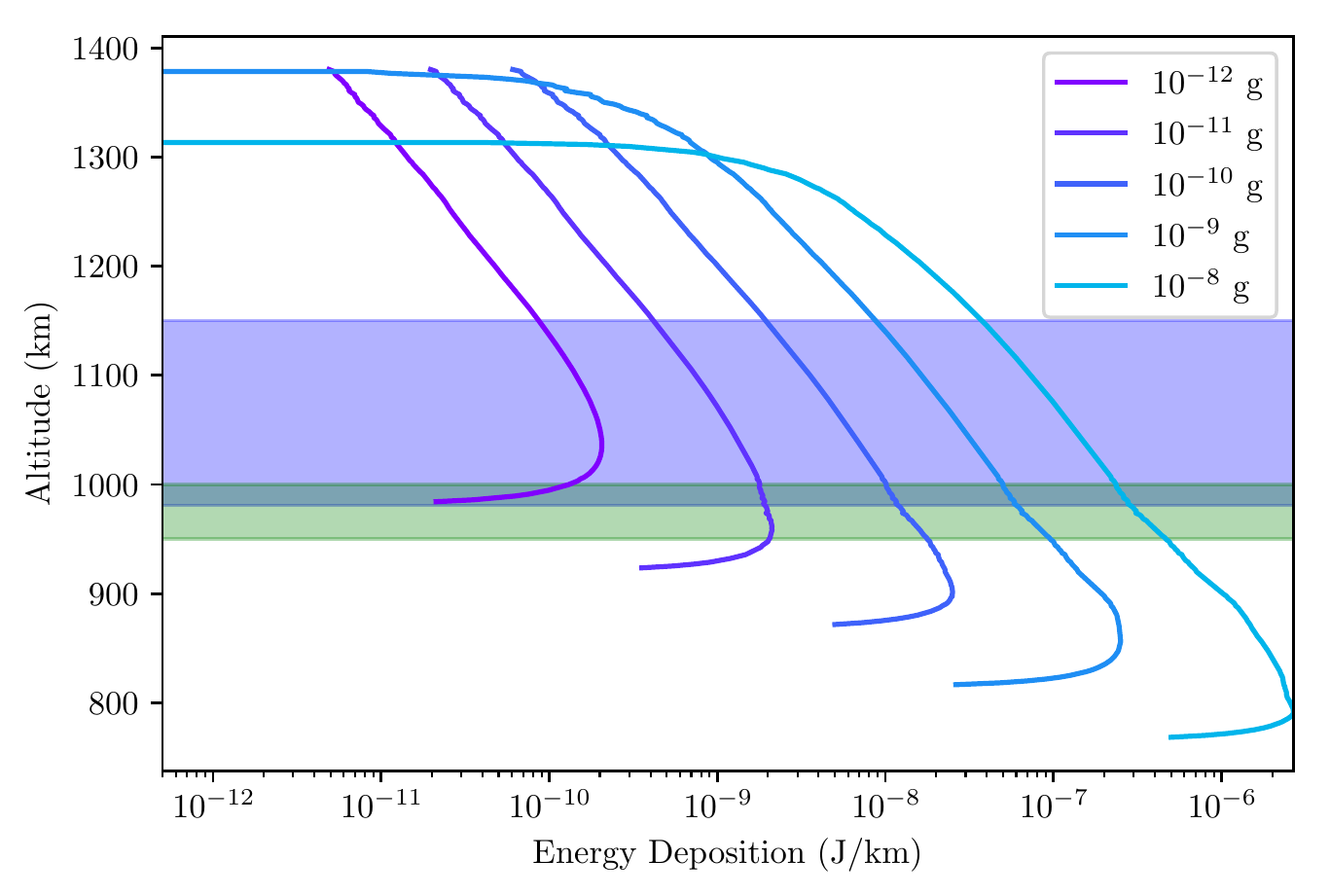}} \subfloat[]{\includegraphics[height=2in, keepaspectratio]{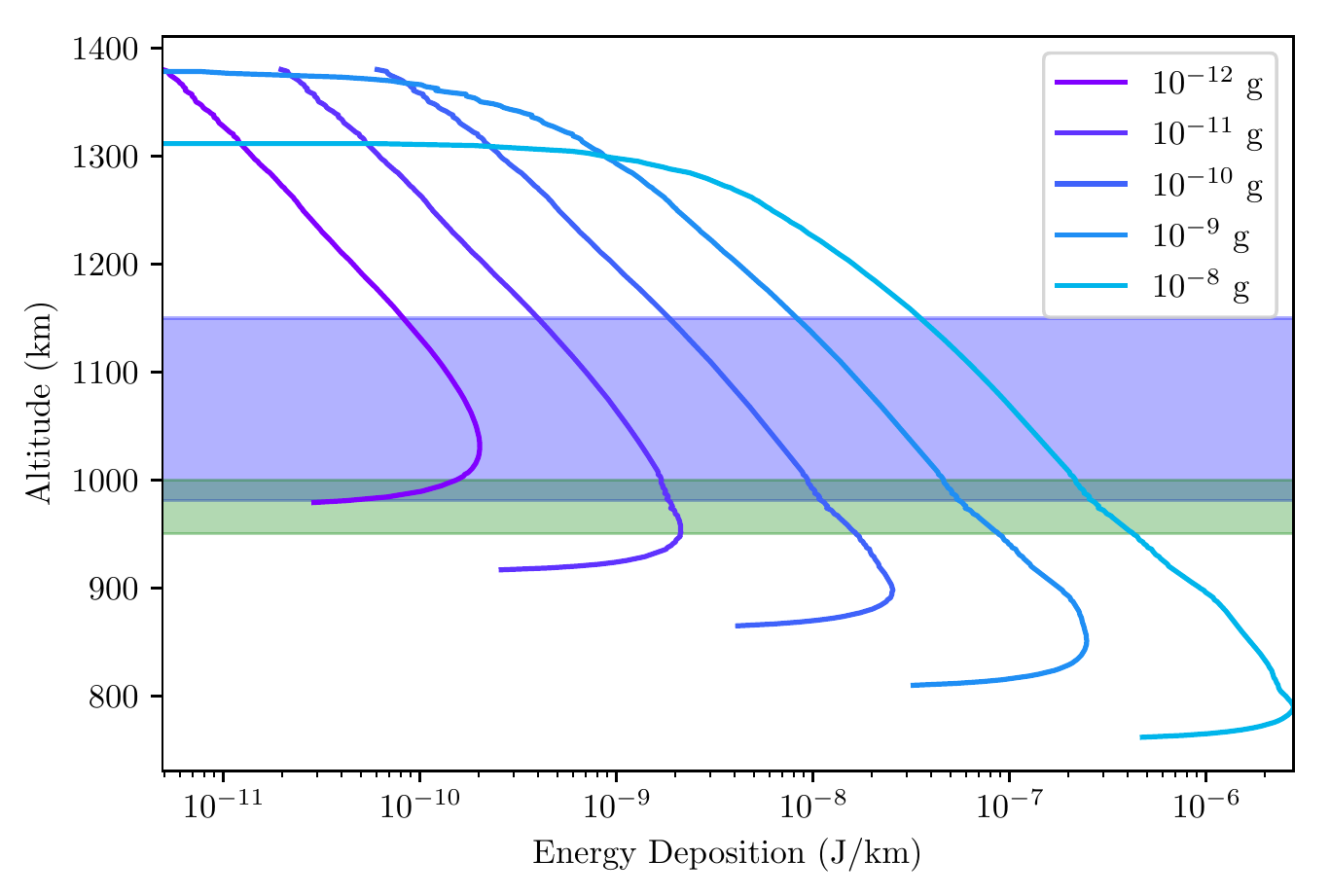}}
    \caption{The shaded regions are the same as in Figure \ref{fig:kin_eng_dep}. For the five smallest mass cases: (a) is the kinetic energy deposition case 1 and (b) is the kinetic energy deposition for case 3}
    \label{fig:smallest_eng}
\end{figure*}

The results for the different cases are plotted in Figures \ref{fig:mass_alt} to \ref{fig:smallest_eng}. Shown first is particle mass as function of altitude for meteors of different initial mass, all incident at the top of the atmosphere at 45$^{\circ}$ and velocity 8 km/s, showing  where in the atmosphere the particles are losing mass and finally being destroyed due to ablation. Following this are figures for energy transfer from the particles' kinetic energy to the atmosphere. The regions of the atmosphere with molecular hydrogen anomalies are shaded green, and the region where there is a methane anomaly is shaded blue in Figure \ref{fig:kin_eng_dep}. These are places where the hydrogen abundance is higher than theoretically predicted, and the methane abundance is lower than theoretically predicted.

First, we note the difference between the models where the trajectory of the particle is and is not calculated, i.e. the upper row of plots versus the lower in Figures 2 and 3. If we just look at Case 1 versus Case 3, for the four smallest particle masses ($10^{-12}$ g, $10^{-11}$ g, $10^{-10}$ g, and $10^{-9}$ g) they survive until a deeper atmospheric depth, depositing energy lower in the atmosphere, when we do calculate the particles' changing trajectory angles, though the effect is minimal.
 
Second, looking at the model where we varied the entry angle and kept the mass the same (Cases 2 and 4), we see little variance in where the particle is destroyed in the atmosphere and depositing energy. The difference is only a couple of tens of kilometers between models. As we anticipated, the 90$^{\circ}$ case does not produce results, as the particle is just glancing off of the top of the atmosphere.

\subsection{Entry Velocity Variations}
\label{sec:vel}
\begin{figure*}
    \centering
    \subfloat[]{\includegraphics[height=2in]{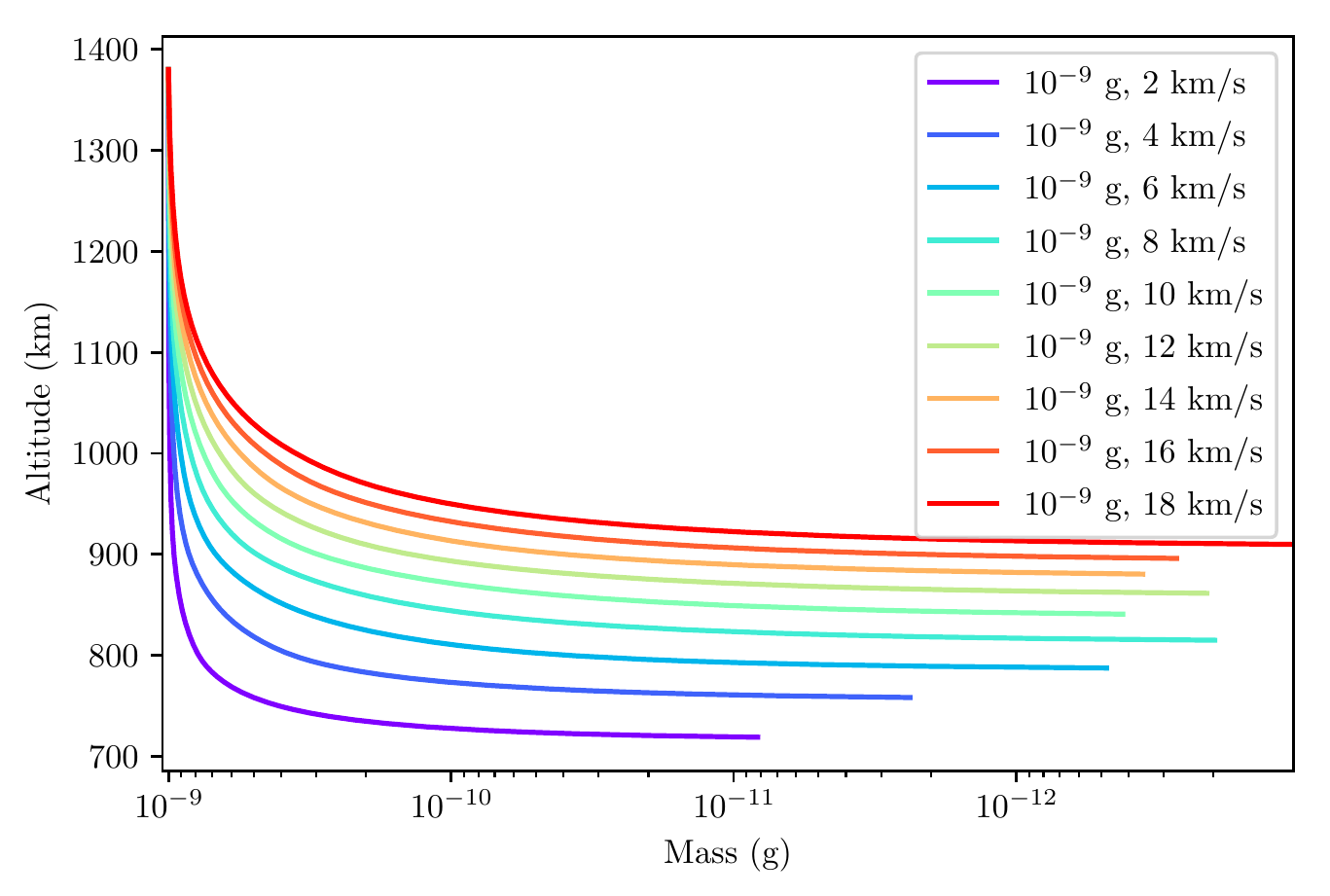}}
    \subfloat[]{\includegraphics[height=2in]{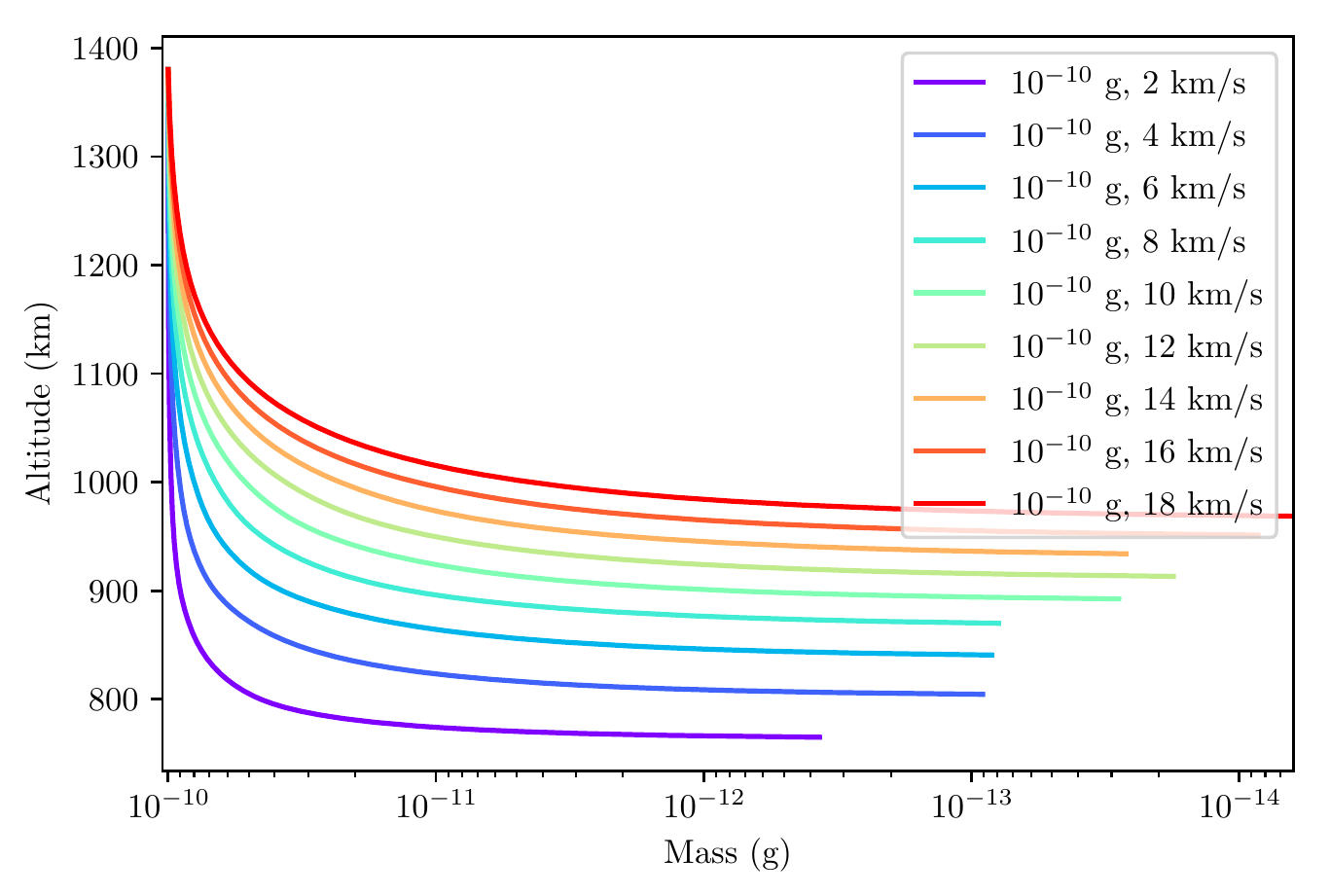}}
    \hspace{0mm}
    \subfloat[]{\includegraphics[height=2in]{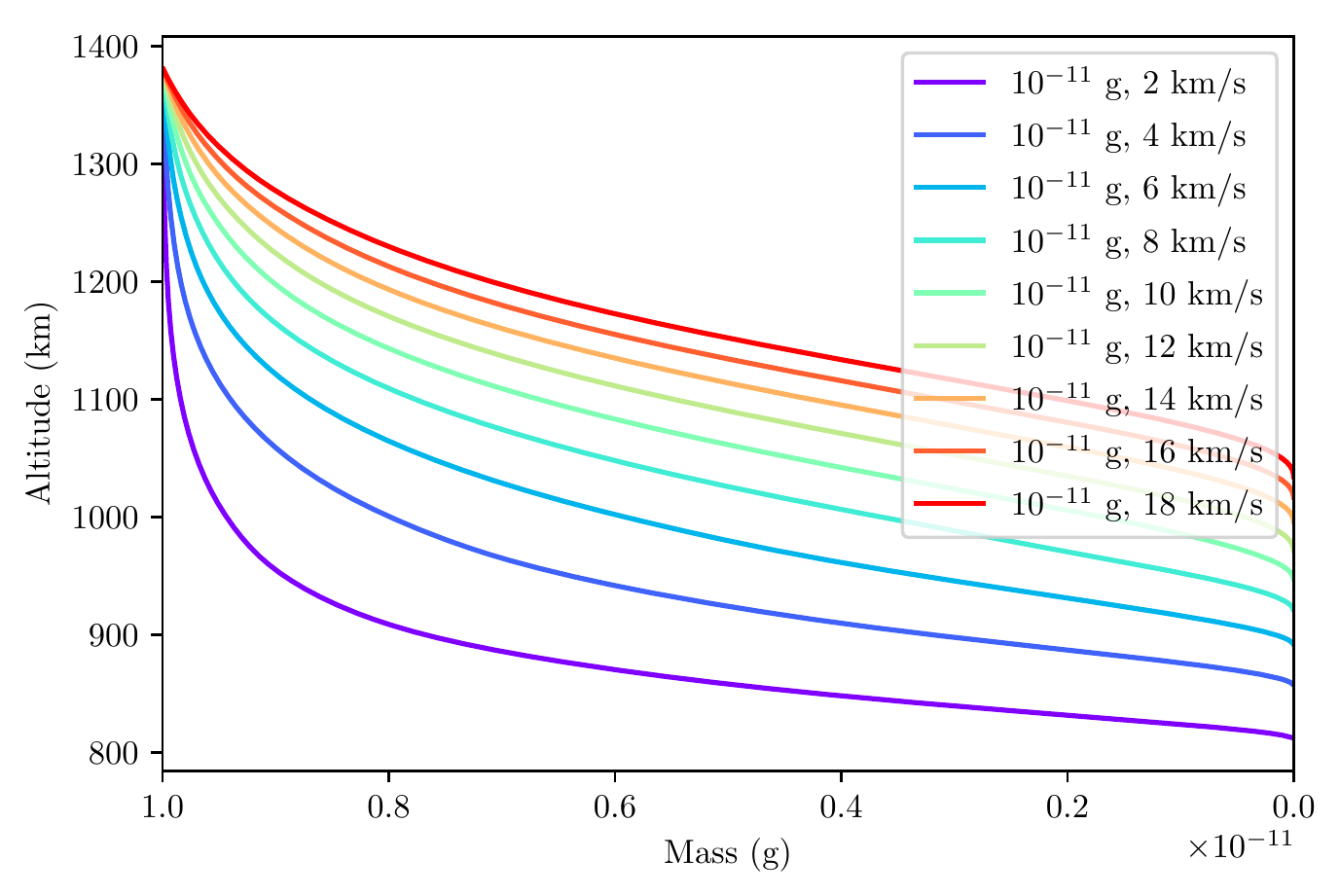}}
    \subfloat[]{\includegraphics[height=2in]{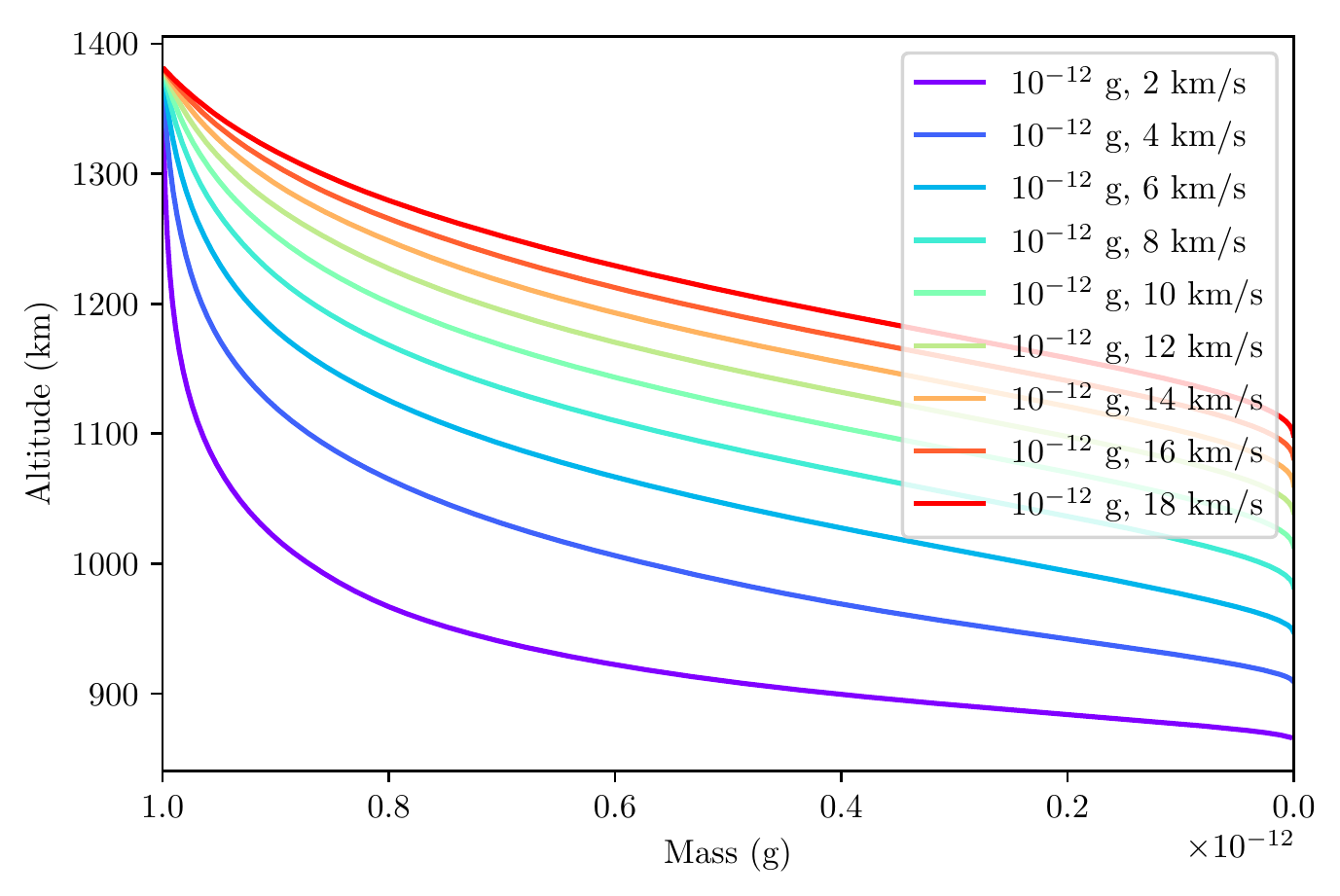}}
    \caption{The mass-altitude plots for entry velocities between 2km/s and 18km/s for the (a) 10$^{-9}$ g, (b) 10$^{-10}$ g, (c) 10$^{-11}$ g, and (d) 10$^{-12}$ g particles. Previous figures all assume that the entry velocity is 8 km/s; here we model how changing the velocity affects particle lifetime and energy deposition. Typical velocities on the leading edge of Titan in its orbit around Saturn will be $\sim 14$ km s$^{-1}$; typical velocities on the trailing edge will be $\sim 2$ km s$^{-1}$, so that typical mass and energy deposition altitude varies by hundreds of kilometers by longitude.}
    \label{fig:vel_test}
\end{figure*}

Titan's orbital velocity around Saturn is $5.6$ km s$^{-1}$. For a typical particle velocity at Titan's distance from Saturn of 8 km $s^{-1}$, this means that velocities on Titan's leading edge will approach 14 km s$^{-1}$, and on its trailing edge will typically be as low as 2 km s$^{-1}$. We therefore tested entry velocities ranging from 2 km/s -- 18km/s in equal intervals for the four smallest particles ($10^{-12}$, $10^{-11}$, $10^{-10}$, and $10^{-9}$ grams) in the case where the angle was held constant at 45$^{\circ}$. We found that in all cases, the altitude at which the particle is destroyed increases with increasing velocity as expected.  The plots of mass versus altitude are shown in Figure \ref{fig:vel_test}. This means that energy and mass deposition on Titan's leading edge in its orbit is typically hundreds of km higher than on its trailing edge. The smallest particles, for common entry velocities, deposit the bulk of their mass and energy in the region of the atmosphere where methane and molecular hydrogen anomalies have been observed.

\section{Discussion}
\label{sec:discuss}
When calculating the total energy available at a given altitude, we use the differential flux values from \cite{Poppe2012EKB} at Titan's orbit. Approximating the grains as spheres, the $10^{-12}$, $10^{-11}$, $10^{-10}$, and $10^{-9}$ gram particles in this study correspond to the 1$\mu$m, 2$\mu$m, 5$\mu$m, and 10$\mu$m in theirs. Taking values for the differential of these particles at Titan, we calculate the total amount of energy in the upper atmosphere over one Saturn-year by multiplying:
\begin{equation}
\label{eq:total_nrg}
\begin{aligned}
    \text{Energy Deposition (J km$^{-1}$) } \times \text{ Flux (km$^{-2}$ s$^{-1}$)} \times 4\pi (R_T + z)^2 \\ \times \ t_S \times \Delta z
\end{aligned}
\end{equation}
where $R_T$ is the radius of Titan in km, $t_S$ is the length of a Saturnian year in seconds, and $z$ is the altitude, $\Delta z$ is the altitudinal step over which energy is deposited (this is $\approx$ 1 km).

\begin{figure*}
    \centering
    \includegraphics[width=\textwidth, keepaspectratio]{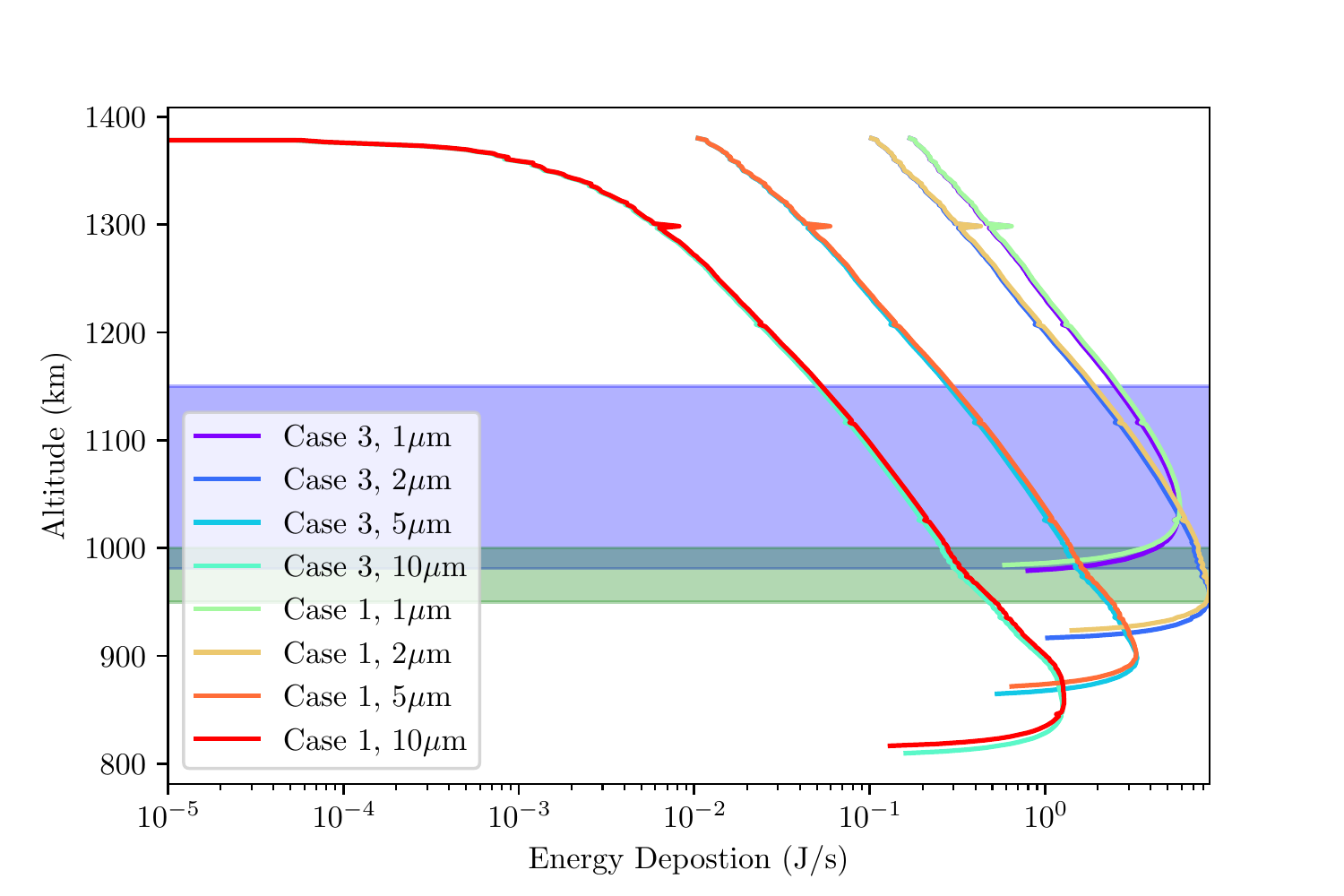}
    \caption{The total amount of energy available per second for the four smallest particle diameters, responsible for the bulk of the mass and energy deposition into Titan's atmosphere.}
    \label{fig:total_nrg}
\end{figure*}

Figure \ref{fig:total_nrg} shows the total energy available at every altitude for the kinetic energy for cases 1 and 3.  In both case 1 and case 3, the 10$^{-11}$ g particle deposits the most energy (2.7$\times 10^{9}$ J), and in case 3, it deposits 2.6$\times 10^{9}$ J when averaging over the entire atmosphere.  Accounting for how $\theta$ changes over the particles' descent does result in the particle being destroyed at lower altitudes, but this difference is small.

\begin{figure*}
    \centering
    \includegraphics[width=\textwidth, keepaspectratio]{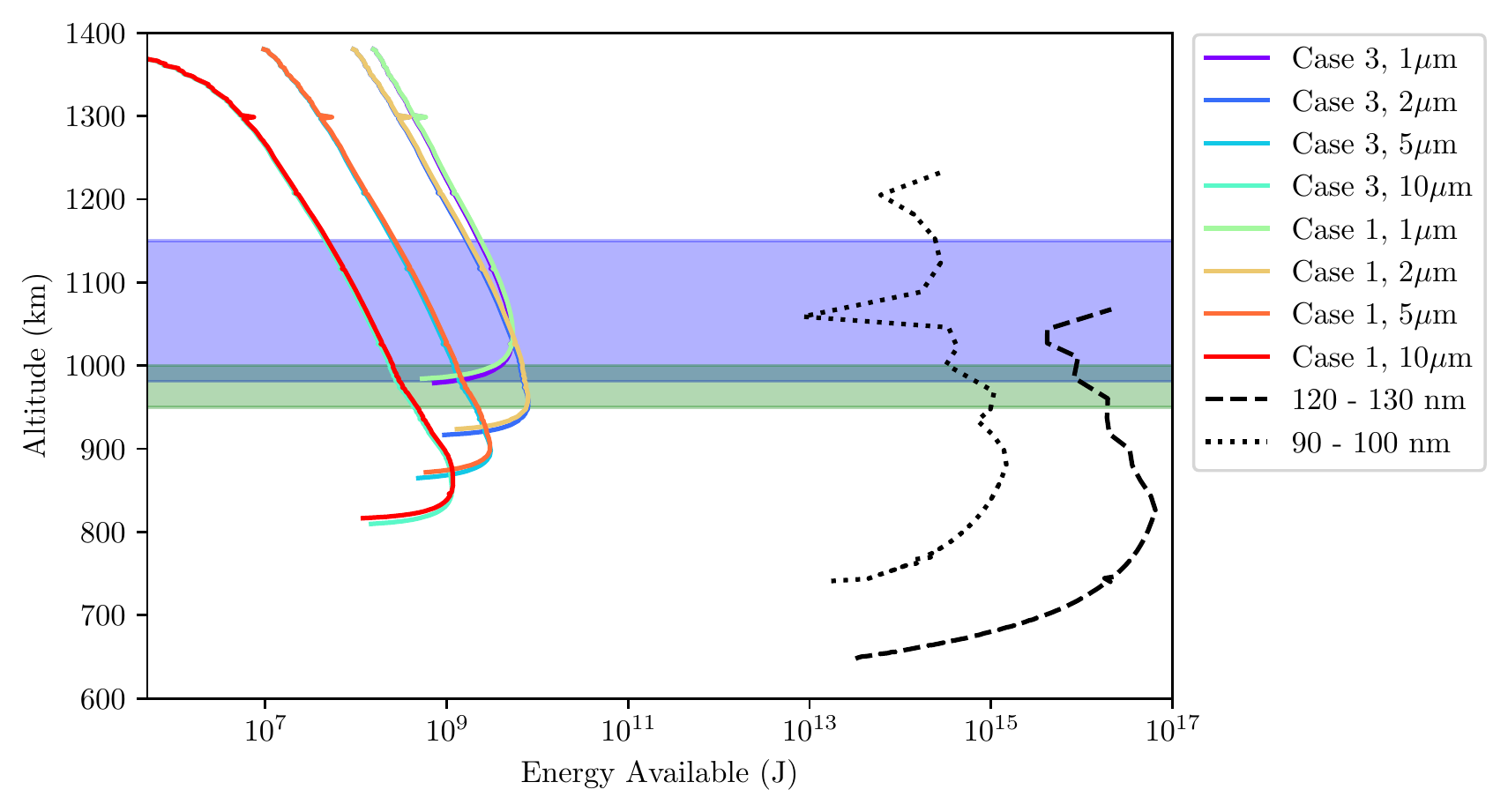}
    \caption{The same as Figure \ref{fig:total_nrg} but with the photon energy, over the course of a Saturnian year, for a comparison of the total energy available from case 1 and case 3 particles, photons in the 120 -- 130 nm range, and photons in the 90 -- 100 nm range. These wavelengths are noted as being the range in which methane photodissociation peaks.}
    \label{fig:total_nrg_comp}
\end{figure*}

We compare this to the energy provided by solar UV photons, which deposit the most energy to Titan's atmosphere. The intensity and absorption rate of those photons varies as a function of altitude (\cite{Krasnopolsky2009Model, Lavvas2009Haze}). Using the absorption curves for 90 -- 100 nm and 120 -- 130 nm from the \cite{Krasnopolsky2009Model} model (CH$_4$ absorption is mostly from 80 -- 140 nm), we compare the total amount of energy from photons available in one Saturnian year at each altitude to the total amount of energy from our meteors. The results, shown in Figure \ref{fig:total_nrg_comp}, confirm that the photons provide the most energy to Titan's atmosphere to drive chemical reactions. While the photons do deposit the most energy when looking over the course of a Saturnian year, it must be noted that the micrometeorites' energy deposition is more important for the night-side of the moon and for the altitudes above $\approx$ 1250 km where the photon energy is not being deposited. Since the micrometeorites are providing both energy and material in the region of the atmosphere above the methane anomaly, they could be a significant driver for chemistry depending on mixing within the atmosphere, though these impacts will be the subject of a follow-up study.

\section{Conclusion}
\label{sec:conclude}
We have tested four different cases of meteors entering Titan's atmosphere and calculated how much energy they are depositing as a function of altitude in order to analyze how they contribute to the atmosphere's overall energy budget. We do this in the interest of trying to understand how various chemical anomalies observed during the Cassini-Huygens mission could possibly have arisen. The four cases were (1) varying the mass while keeping a constant angle at every altitude step, (2) varying the entry angle while keeping a constant angle at every altitude step, (3) varying the mass and varying the meteorite's angle relative to horizontal at every altitude step, and (4) varying the entry angle and varying the meteorite's angle relative to horizontal at every altitude step. We found that in varying the angle throughout the lifetime of the meteorite results in its evaporation higher in the atmosphere. In the case of the smallest particles which make up the bulk of the material falling onto Titan, the altitudes at which they are destroyed and depositing energy are the altitudes wherein anomalies in the methane and molecular hydrogen mixing ratios are observed. 

When calculating the total amount of energy provided by the smallest IDPs, using fluxes from \cite{Poppe2012EKB}, energy is deposited slightly deeper in the atmosphere at many of the same altitudes when one accounts for changes in trajectory over the course of the particles' lifetimes compared to when we don't allow for changes in trajectory. The 10$^{-11}$ g particles deposit the most energy over a Saturnian year, supplying the upper atmosphere with $2.7\times10^{9}$ J of energy (case 1)/$2.6\times10^{9}$ (case 3). Although this is small compared to the total energy provided by solar UV photons, it is of course the dominant energy source during the Titanian night. Further study is required to see if the amount of carbon, hydrogen, oxygen, and other key elements delivered by micrometeorites could potentially explain, at least in part, the anomalies observed in the upper part of Titan's atmosphere.

\clearpage
\bibliographystyle{cas-model2-names}
\balance
\bibliography{bibliography}

\end{document}